\documentclass[prl,superscriptaddress,a4paper,twocolumn]{revtex4-1}
\usepackage{geometry}
\usepackage[english]{babel}
\usepackage[latin1]{inputenc}
\usepackage{hyperref}
\usepackage{amsmath, amssymb, amsfonts, mathrsfs}
\usepackage{graphicx}

\graphicspath{{images/}}
\usepackage{xcolor}
\usepackage{exscale}
\usepackage{graphicx}
\usepackage{amsmath}
\usepackage{latexsym}
\usepackage{amsfonts}
\usepackage{amssymb}
\usepackage{bbm}

\def\pmx{\begin{pmatrix}}
\def\emx{\end{pmatrix}}

\newcommand{\ket}[1]{|#1\rangle}

\begin{document} 

\title{Entanglement by Path Identity}

\author{Mario Krenn}
\email{mario.krenn@univie.ac.at}
\affiliation{Vienna Center for Quantum Science \& Technology (VCQ), Faculty of Physics, University of Vienna, Boltzmanngasse 5, 1090 Vienna, Austria.}
\affiliation{Institute for Quantum Optics and Quantum Information (IQOQI), Austrian Academy of Sciences, Boltzmanngasse 3, 1090 Vienna, Austria.}
\author{Armin Hochrainer}
\affiliation{Vienna Center for Quantum Science \& Technology (VCQ), Faculty of Physics, University of Vienna, Boltzmanngasse 5, 1090 Vienna, Austria.}
\affiliation{Institute for Quantum Optics and Quantum Information (IQOQI), Austrian Academy of Sciences, Boltzmanngasse 3, 1090 Vienna, Austria.}
\author{Mayukh Lahiri}
\affiliation{Vienna Center for Quantum Science \& Technology (VCQ), Faculty of Physics, University of Vienna, Boltzmanngasse 5, 1090 Vienna, Austria.}
\affiliation{Institute for Quantum Optics and Quantum Information (IQOQI), Austrian Academy of Sciences, Boltzmanngasse 3, 1090 Vienna, Austria.}
\author{Anton Zeilinger}
\email{anton.zeilinger@univie.ac.at}
\affiliation{Vienna Center for Quantum Science \& Technology (VCQ), Faculty of Physics, University of Vienna, Boltzmanngasse 5, 1090 Vienna, Austria.}
\affiliation{Institute for Quantum Optics and Quantum Information (IQOQI), Austrian Academy of Sciences, Boltzmanngasse 3, 1090 Vienna, Austria.}

\begin{abstract}
Quantum entanglement is one of the most prominent features of quantum mechanics and forms the basis of quantum information technologies. Here we present a novel method for the creation of quantum entanglement in multipartite and high-dimensional systems. The two ingredients are 1) superposition of photon pairs with different origins and 2) aligning photons such that their paths are identical. We explain the experimentally feasible creation of various classes of multiphoton entanglement encoded in polarization as well as in high-dimensional Hilbert spaces -- starting only from non-entangled photon pairs. For two photons, arbitrary high-dimensional entanglement can be created. The idea of generating entanglement by path identity could also apply to other quantum entities than photons. We discovered the technique by analyzing the output of a computer algorithm. This shows that computer designed quantum experiments can be inspirations for new techniques.
\end{abstract}

\maketitle

\begin{figure}[ht!]
\includegraphics[width=0.5 \textwidth]{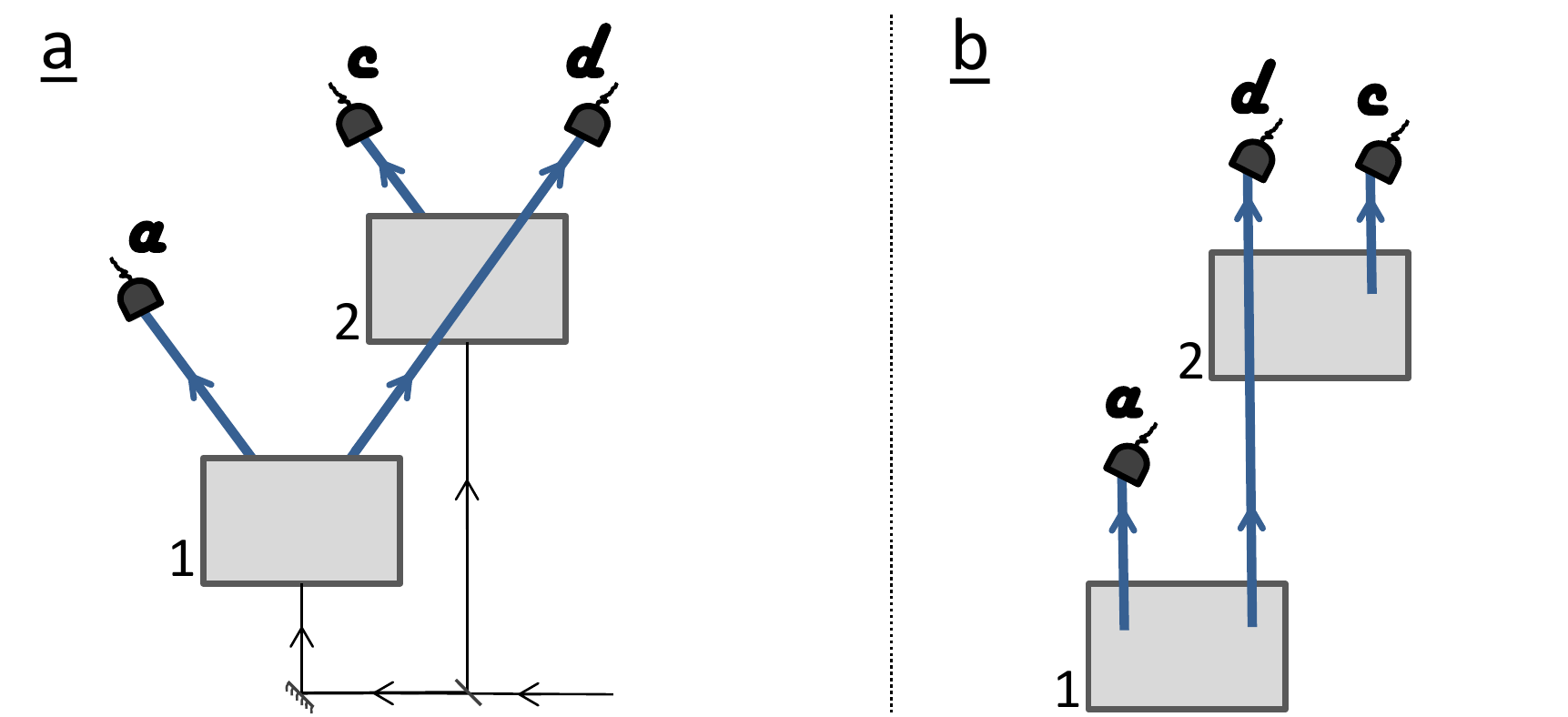}
\caption{\textbf{a}: The simplest example which uses the overlapping modes has been discussed first in \cite{wang1991induced}. The two crystals (grey squares) can produce one pair of photons (blue lines), either in the first or in the second crystal with the pump beam depicted with black lines. If the two processes are coherent and the photons have same the frequency and polarization, it is not known in which crystal the photons are created. In that case, the resulting photon pair is in the state $\ket{\psi}=\frac{1}{\sqrt{2}}\left(\ket{a}+\ket{c}\right)\ket{d}$. \textbf{b}: A simple sketch of the same experiment. For simplicity, we will use this more abstract representation of physical experiments in the rest of the manuscript.}  
\label{fig:Mandel}
\end{figure}

In 1991 Zou, Wang and Mandel reported an experiment where they induce coherence between two photonic beams without interacting with any of them \cite{wang1991induced, zou1991induced}. They used two SPDC (spontaneous parametric down-conversion) crystals, where one photon pair is in a superposition of being created in crystal 1 and crystal 2 -- which can be described as $\ket{\psi}=\frac{1}{\sqrt{2}}\left(\ket{a}\ket{b}+\ket{c}\ket{d}\right)$. The striking idea (originally proposed by Zhe-Yu Ou) was to overlap one of the paths from each crystal (Figure \ref{fig:Mandel}), which can be written as $\ket{b}=\ket{d}$. This method removes the \textit{which-crystal information} of the final photon in path $d$. In contrast to a quantum eraser, the information here is not erased by postselection. Instead, all photons arrive in the same output irrespectively in which crystal they are created. The resulting state can be written as $\ket{\psi}=\frac{1}{\sqrt{2}}\left(\ket{a}+\ket{c}\right)\ket{d}$: One photon is in path $d$, and its partner photon is in a superposition of being in path $a$ and $c$. There has been some follow-up work in recent years in the areas of quantum spectroscopy \cite{kulik2004two, kalashnikov2016infrared}, quantum imaging \cite{lemos2014quantum, lahiri2015theory}, studies of complementarity \cite{herzog1994frustrated, herzog1995complementarity, heuer2015induced, heuer2015complementarity}, optical polarization \cite{lahiri2015partial} and in microwave superconducting cavities \cite{lahteenmaki2016coherence}. However, this striking idea has not been investigated in the context of quantum entanglement generation yet.

Here we show that by superposing photon pairs created in different crystals, and overlapping the photons paths, one can generate very flexible experiments producing various types of entanglement, both in the multiphoton and the high-dimensional regime. We start by presenting different schemes to produce various multiphoton polarization-entangled states such as Greenberger-Horne-Zeilinger (GHZ) states \cite{greenberger1989going} and W states \cite{zeilinger1997nasa, bourennane2004experimental}, and contrast them with traditional methods of creating these states \cite{pan2012multiphoton}. The method is then generalized to high-dimensional multiphoton entangled states (such as a 4-particle 3-dimensional GHZ state), which so far can only be produced in a few special cases \cite{malik2016multi, krenn2016automated}. Furthermore, we present for the first time a method to create arbitrary high-dimensional two-photon entangled states, for example, arbitrary high-dimensional Bell states in orbital angular momentum (OAM) or frequency of photons.
\begin{figure}[t]
\includegraphics[width=0.5 \textwidth]{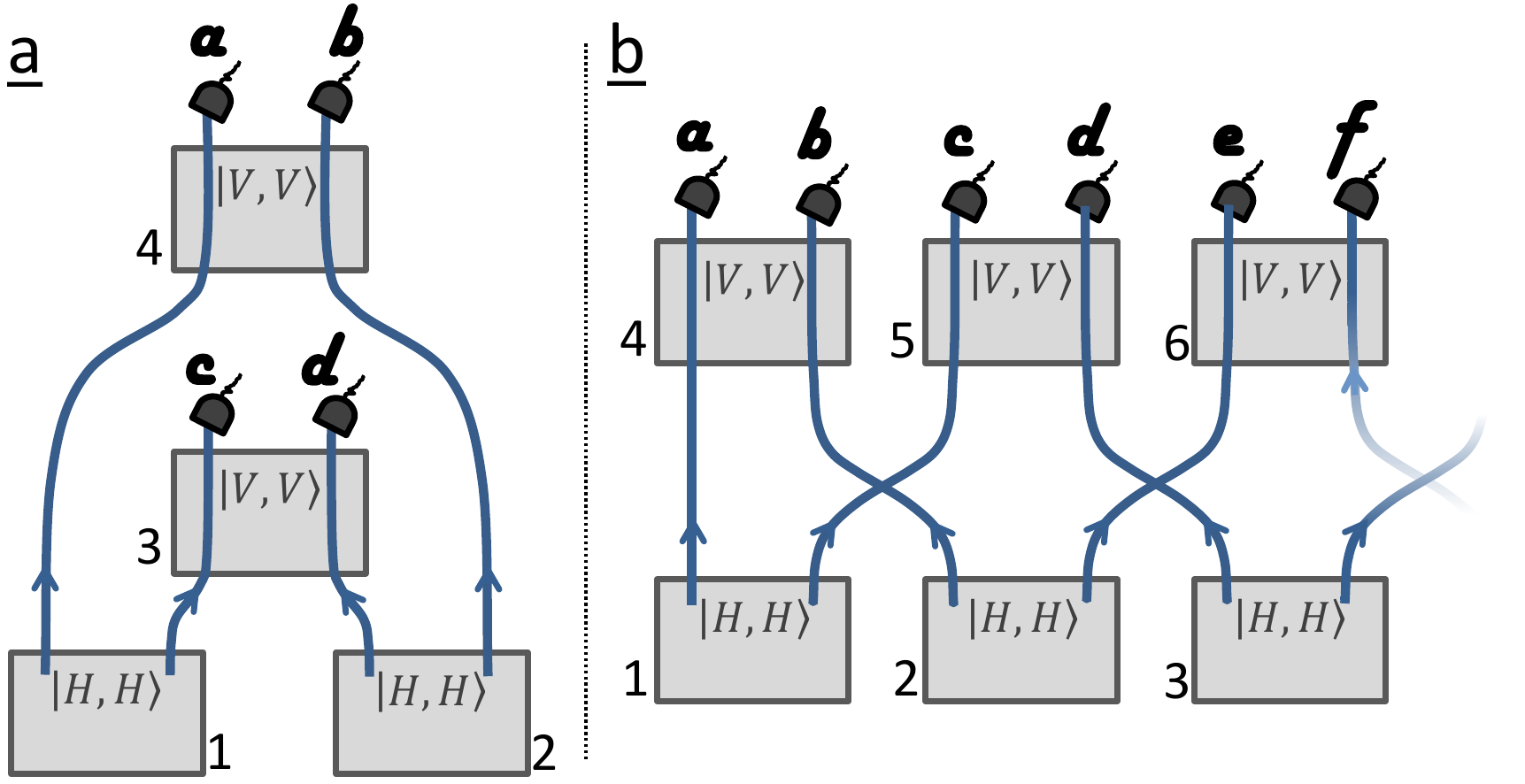}
\caption{Multiphoton entanglement with polarization. \textbf{a}: In four crystals, two photon-pairs are produced. Crystal 1 and 2 produce horizontally polarized photons, while crystal 3 and 4 produce vertically polarized ones. Four-photon coincidences can only happen when crystal 1 and 2 fire together or when crystal 3 and 4 fires. This leads to a 4-particle GHZ-state $\ket{\psi}=\frac{1}{\sqrt{2}}\left(\ket{H,H,H,H}+\ket{V,V,V,V}\right)$. \textbf{b}: Entangled states with more numbers of particles can be created in an analogous way -- here a $n$-photon GHZ state $\ket{\psi}=\frac{1}{\sqrt{2}}\left(\ket{H,H,H,...}+\ket{V,V,V,...}\right)$ is shown.}  
\label{fig:4GHZpol}
\end{figure}

\textit{Multi-photon entanglement in Polarization} -- First, we consider 4-photon polarization entanglement (Figure \ref{fig:4GHZpol}a). Crystal 1 and 2 can produce horizontally polarized pairs while crystal 3 and 4 can produce vertically polarized ones. The crystals are pumped coherently and the pump power is adjusted such that we can neglect the cases where more than two pairs are created. The idea is that four-photon coincidences (i.e. one photon in each of the four paths) can only happen either when the two pairs come from crystal 1 \& 2 or crystal 3 \& 4. No other event produces four-photon coincidences. For example, if the pairs are produced in crystal 1 \& 3, there will be two photons in path $c$, but none in path $b$. The resulting four-photon state can be written as (see Appendix for a detailed calculation)
\begin{align}
\ket{\psi}=\Big(\ket{H_a,H_c}+\ket{H_b,H_d}+\ket{V_a,V_b}+\ket{V_c,V_d}\Big)^2,\nonumber\\
\ket{\psi} \xrightarrow{\text{4-fold}} \frac{1}{\sqrt{2}}\Big(\ket{H_a,H_b,H_c,H_d}+\ket{V_a,V_b,V_c,V_d}\Big),
\label{4partyGHZ}
\end{align}
where $H$ and $V$ stand for horizontal and vertical polarization, respectivly, and the subscript stands for the photon's path. The final result is a 4-photon GHZ-state. A realistic diagram of the experimental setup as well as discussion about requirements for temporal coherence and indistinguishability (applying the methods from \cite{jha2008temporal} to the four-photon case) can be found in the Appendix. In an analogous way, by increasing the number of crystals and the pump power, entangled states with more photons can be created. In figure \ref{fig:4GHZpol}b the scheme for creating a n-photon GHZ state is shown.
\begin{figure}[ht!]
\includegraphics[width=0.5 \textwidth]{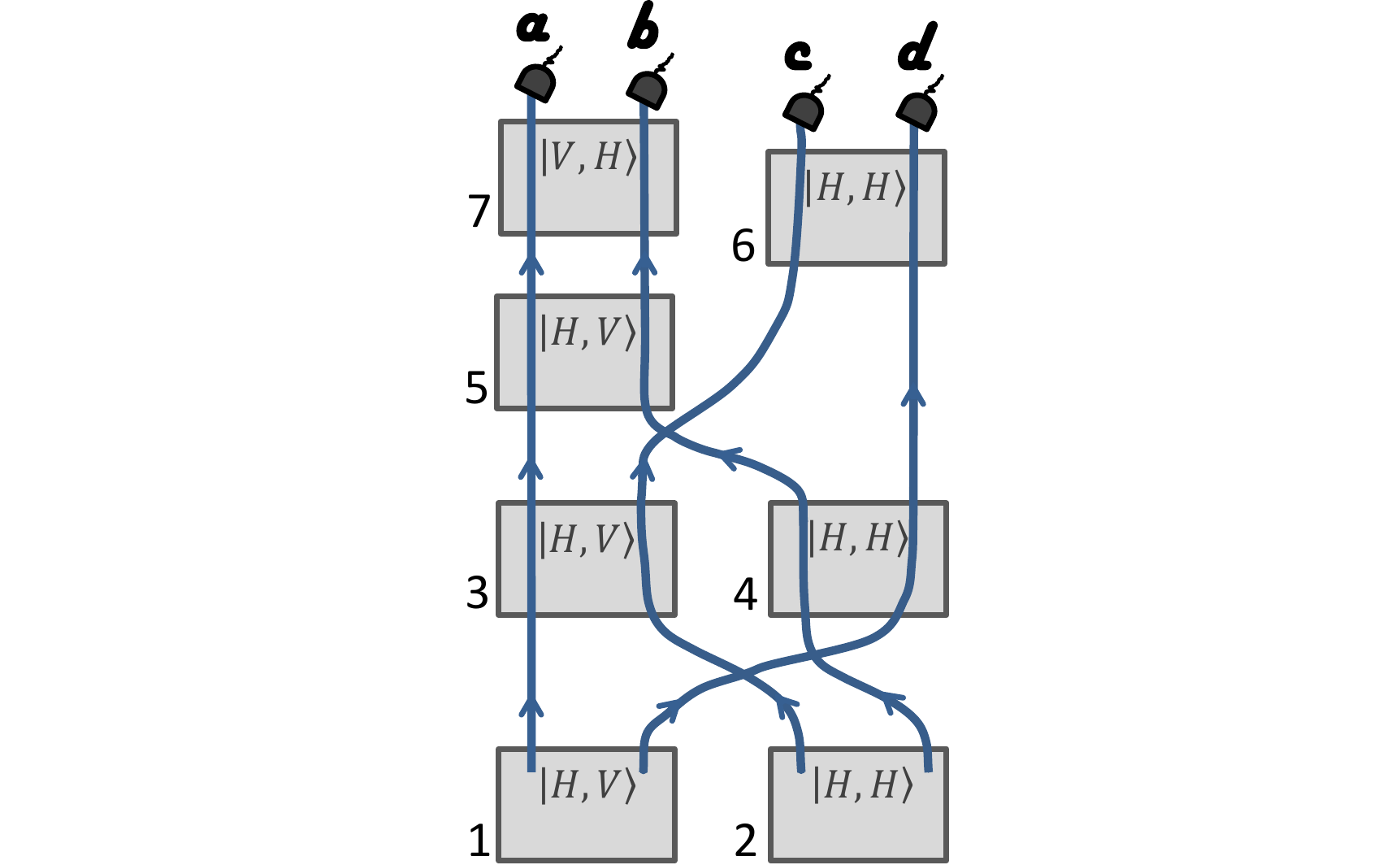}
\caption{W-states such as $\ket{\psi}=\frac{1}{2}\left(\ket{VHHH}+\ket{HVHH}+\ket{HHVH}+\ket{HHHV}\right)$ represent a different type of multi-photon entanglement. While the GHZ state is considered as the most non-classical state, a W-state is the most robust entangled state because the loss of one particle leaves an entangled state. Here, four-folds can only happen if crystal 1 \& 2 produces both a pair of photons, or crystal 3 \& 4, or crystal 5 \& 6 or 6 \& 7. Interestingly, in this setup, one photon from crystal 1 can stimulate an emission in crystal 3. However, this will not lead to four-fold coincidences, as there is no photon in path $b$. Thus, this event can be neglected.}  
\label{fig:4Wpol}
\end{figure}

In contrast to our new method, the traditional way of creating 4-photon entangled states requires two crystals each producing a pair of polarization entangled photons. One photon from each crystal goes to a polarizing beam splitter (PBS), which removes the which-crystal information. Triggering on events where all 4 detectors click, a 4-particle GHZ $\ket{\psi}=\frac{1}{\sqrt{2}}\left(\ket{H,H,H,H}+\ket{V,V,V,V}\right)$ state is created \cite{pan2001experimental}. With that traditional method, GHZ-entanglement with 8 photons \cite{yao2012observation, huang2011experimental} and very recently, up to 10 photons have been created \cite{wang2016experimental, chen2017observation}. 

The new scheme does not need entangled photons to start with. Furthermore removing of the which-crystal-information using a PBS is not necessary, as it has never been created in the first place. Our method does not use cascaded down-conversion, as it has been shown in recent articles producing multiphoton polarization entangled states \cite{hubel2010direct, hamel2014direct}. In our examples, stimulated emission does not happen (which would introduce noise in the entangled state), because the input modes into the crystal are orthogonal to the output modes (for instance, having different polarization). In other cases, such as for the generation of the 4 photon W-state (figure \ref{fig:4Wpol}), stimulated emission can happen but its contributions to the four-photon coincidences are negligible.
\begin{figure}[ht!]
\includegraphics[width=0.5 \textwidth]{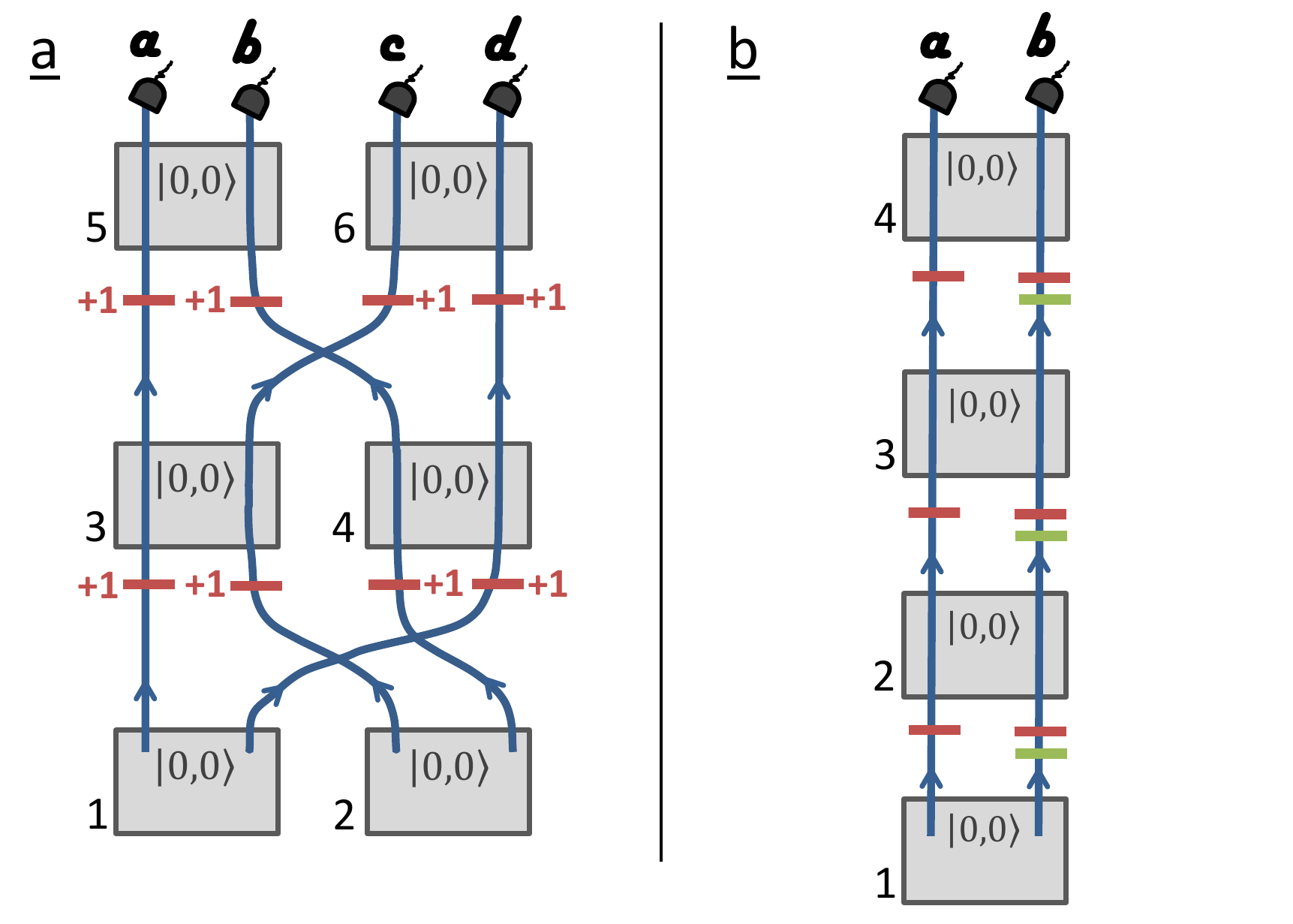}
\caption{Multiphoton-Entanglement with high-dimensional degrees of freedom (in this example: orbital angular momentum). \textbf{a}: The setup produces two photon-pairs. Four-Photon coincidences can only occur when crystal 1 \& 2 fire together, or crystal 3 \& 4 or crystal 5 \& 6. The produced photon pairs are all in the lowest mode (such as OAM=0). After each layer of crystals, a hologram increases the OAM of the photons (depicted as red line). After the third layer, photons from crystal 1 \& 2 have OAM=2 and photons from the middle layer have OAM=1, which leads to a four-particle three-dimensional entangled GHZ-state ($\ket{\psi}=\frac{1}{\sqrt{3}}\left(\ket{0,0,0,0}+\ket{1,1,1,1}+\ket{2,2,2,2}\right)$). \textbf{b}: The same technique can also be applied to two-particle states to produce general high-dimensional states. All crystals produce pairs of Gaussian photons. The red lines indicate OAM holograms, the green lines indicate phase shifters. For example, if the holograms all are OAM=1 and phases are ignored, then the final output state is $\ket{\psi}=\frac{1}{2}\left(\ket{0,0}+\ket{1,1}+\ket{2,2}+\ket{3,3}\right)$. However, if one changes phases and the OAM in photon b, arbitrary 4-dimensional states can be created -- for example all 16 four-dimensional Bell states. By increasing the number of crystals, more dimensions can be added.}  
\label{fig:4partyhighdim}
\end{figure}

\textit{Multi-photon entanglement in higher dimensions} -- The principle can be generalized to produce high-dimensional multiphoton entangled states \cite{huber2013structure, lawrence2014rotational}. High-dimensional entanglement has been investigated mainly in the two-photon case \cite{dada2011experimental, romero2012increasing, krenn2014generation, zhang2016engineering}, with two recent exceptions which investigated three-dimensional entanglement with three photons \cite{malik2016multi}, and teleportation of two degrees-of-freedom of a single photon \cite{wang2015quantum}. Figure \ref{fig:4partyhighdim}a shows our proposal for an experiment creating a 3-dimensional 4-party GHZ-state, starting from crystals which create separable photon pairs. There are 3 layers of 2 crystals -- i.e. 6 crystals that are pumped coherently, and photon pairs are created in two of them (because the pump power is set to such a level that higher-order emissions can be neglected). 

Each photon from the first layer (crystals 1 and 2) passes through two mode shifters, which in total shift its mode by +2. In the case of orbital angular momentum (OAM) of photons \cite{krenn2017orbital}, mode shifters are holograms which add one unit of OAM to the photon (an analogous method could be done with discrete frequency or time bins \cite{richart2012experimental, martin2017quantifying}). Photons from the second layer (crystal 3 and 4) pass through one mode shifter, while the photons created in the uppermost layer (crystal 5 and 6) stay in their initial mode. The resulting 4-photon state can be described in the following way (see Appendix for details):
\begin{align}
\ket{\psi}=\Big(\ket{2_a,2_d}+\ket{2_b,2_c}+\ket{1_a,1_c}+\ket{1_b,1_d}\nonumber\\
+\ket{0_a,0_b}+\ket{0_c,0_d}\Big)^2,
\label{highdimsqare}
\end{align}
where $0$, $1$ and $2$ stand for the mode number (such as the OAM of the photon), and the subscript denotes the photon's path. In the same way as before, by neglecting cases where more than two photon pairs are produced, one finds that a 4-fold coincidence event in detectors a, b, c, d can only be created either if crystal 1 \& 2 fire together or crystal 3 \& 4 or crystal 5 s\& 6. This leads to
\begin{align}
\ket{\psi}\xrightarrow{\text{4-fold}} \frac{1}{\sqrt{3}}\Big(\ket{2_a,2_b,2_c,2_d}+\ket{1_a,1_b,1_c,1_d}\nonumber\\
+\ket{0_a,0_b,0_c,0_d}\Big).
\label{4party3Dim}
\end{align}
Our idea can further be generalized to cases of more than 4 photons. As an example, a scheme for 6 photons entangled in 5 dimensions is shown in the Appendix. In general, adding additional columns (and increasing the pump power) increases the photon number $n$, while adding additional layers increases the dimensionality of entanglement $d$. That allows for the creation of arbitrary $n$-photon states entangled in $d=n-1$ dimensions dimensions (for even $n$, and $d=n$ for odd $n$ when one photon is used as a trigger). Furthermore, symmetries of the states can be exploited which lead to a vast number of available entangled states (such as asymmetrically entangled states, which exist only when $n$ and $d$ are both larger than two \cite{huber2013structure}). A detailed analysis of which states can be produced in this way is given in the in Appendix. 

The efficiency, $E$, of state generation is the probability of getting a desired state from all n-fold photon terms. The GHZ state has an efficiency of $E=\frac{d}{\left(\frac{n\cdot d}{2} \right)^{n/2}}$, other states have higher efficiencies (details in Appendix). The efficiency of this method and the commonly used technique for polarization GHZ states are the same \cite{wang2016experimental}. The expected efficiency of a 3-dimensional 3-photon GHZ state (the only high-dimensional GHZ state where the experimental implementation is known) with our new technique is significantly higher than the known technique \cite{krenn2016automated}.

\textit{Two-photon arbitrary high-dimensional entanglement} -- Finally, we show that the same technique can be applied to generate arbitrary high-dimensional entangled two-photon states, starting again with only separable (non-entangeld) photon pairs. As shown in figure \ref{fig:4partyhighdim}b, four crystals are set up in sequence (only one photon pair is produced) and their output modes are overlapped. Between each crystal, one adds arbitrary phase shifters and mode shifters. That allows for adjusting every individual term in the superposition independently. For example, with all phases set to $\phi=\frac{\pi}{2}$ and all mode shifters being +1, the setup creates $\ket{\psi}=\frac{1}{2}\left(\ket{0,0}+i\ket{1,1}-\ket{2,2}-i\ket{3,3}\right)$. The dimension can be increased by increasing the number of layers (crystals); the minimum number of layers for creating a $d$-dimensional entangled state is $d$.

Traditional methods for producing high-dimensional entanglement exploit the entanglement produced directly in a crystal. Such methods can only produce very restricted type of states. Furthermore, those states are never maximally entangled and have low rates of production. Our technique overcomes these restrictions and can produce arbitrary high-dimensionally entangled two-photon states. We can also tune the amount of entanglement in the following ways: 1) by adjusting the pump laser power between different crystals, we can produce non-maximally entangled pure states; 2) by pumping the crystals with pumps that are not fully coherent to each other, we can produce entangled mixed states.

The number of photon pairs created does not depend on the number of crystals in the experimental setup. For example in figure \ref{fig:4partyhighdim}b, even though there are 4 crystals, only one photon pairs are created. Therefore the expected two-photon rate is of the same order as in a conventional single-crystal source. Moreover, our method requires only separable photon pairs to begin with. Therefore, for OAM of photons, the production rates can be significantly higher than the rate achievable with a traditional method (where higher-dimensional entanglement created directly in the crystal). This is because it is substantially easier to create photon pairs in zero-order (Gaussian spatial mode) than in higher-order modes.

Interestingly, the simplest special case of the technique presented here is a commonly used source of two-photon polarization-entanglement. The so-called \textit{cross-crystal source} uses two crystal after each other, where the first one can create a horizontally polarized photon pair, the second one creates vertically polarized photon pairs \cite{kwiat1999ultrabright}. Pumping both crystal at the same time and producing one pair of photons, one can create a $\ket{\psi}=\frac{1}{\sqrt{2}}\left(\ket{H,H}+\ket{V,V}\right)$ state. That technique can now be seen as a special case of a much broader technique to produce highly flexible high-dimensional multiphoton states in various degrees of freedom, by exploiting superposition of photon-pair origins and overlapping of paths of photons.

To conclude, we investigate new types of photonic entanglement generation by combining two methods: First, photon pairs which originate from different crystals are coherently superposed. Second, we align photon paths to manipulate the structure of the entangled state. It allows for the generation of very general quantum entangled states, for high-dimensional and multi-photonic systems. Our method can be favorably suited for photonic quantum computation schemes particularly in miniaturized compact devices.

Topics for future research involve: Exploring the relation to generating entanglement by propagation, detection and post selection \cite{thiel2007generation}, by using the indistinguishability \cite{killoran2014extracting, sasaki2011entanglement, tichy2013limits} or by using linear optics \cite{migdal2014multiphoton}; Treating the temporal coherence (such as investigated in \cite{jha2008temporal} for two-photon states) in general for multi-photon experiments; Generalizing the creation of entanglement by path identity (or more generally, identity of \textit{some} degree of freedom) to other quantum entities, e.g. microwave superconducting cavities \cite{lahteenmaki2016coherence}, atomic systems \cite{keller2014bose, lopes2015atomic}, trapped ions \cite{blatt2008entangled}, superconducting circuits \cite{barends2014superconducting}.

Finally, we discovered this technique by analyzing the output of a computer algorithm which designs new quantum optical experiments \cite{krenn2016automated}. From there, we generalized the idea (see Appendix). It shows that automated designs of quantum optical experiments by algorithms can not only produce specific quantum states or transformations, but can also be a source of inspiration for new techniques -- which can further be investigated by human scientists. 

\section*{Acknowlegdements}
The authors thank Melvin for investigating 100 million quantum optical experiments. We thank Manuel Erhard and Thomas Scheidl for helpful discussions. We also thank Dominik Leitner for providing computation resources. This work was supported by the Austrian Academy of Sciences (\"OAW), by the European Research Council (SIQS Grant No. 600645 EU-FP7-ICT) and the Austrian Science Fund (FWF) with SFB F40 (FOQUS) and FWF project CoQuS No. W1210-N16.
\bibliographystyle{unsrt}
\bibliography{refs}

\section{Appendix I. Realistic diagram of the polarization GHZ setup and coherence time requirements}\label{app1}
\begin{figure}[ht!]
\includegraphics[width=0.5 \textwidth]{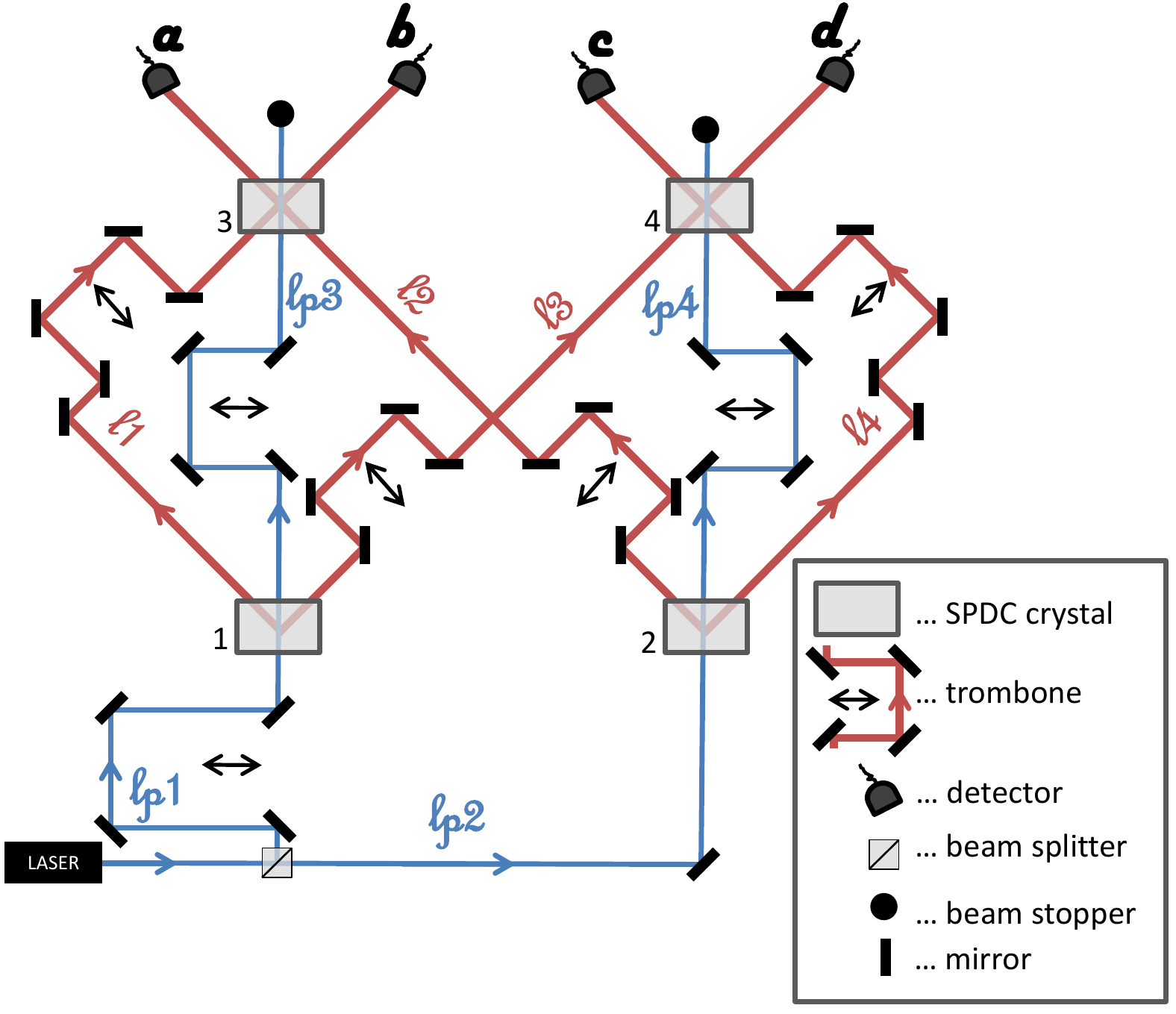}
\caption{A realistic diagram of the experiment described in the main text, which produces a 4-photon GHZ-state in polarization. The pump length $\ell_{p1}$ and $\ell_{p2}$ start at the BS and end at crystals 1 and 2. The lengths $\ell_{p3}$ and $\ell_{p4}$ start at crystals 1 and 2, and end at crystals 3 and 4. The lengths for the down-conversion photons $\ell_i$ start at the crystals 1 and 2, and end at crystals 3 and 4.} 
\label{figSI:4GHZ}
\end{figure}
A possible implementation of the scheme described in figure \ref{fig:4GHZpol}a , which produces a 4-photon GHZ state in polarization, is shown in more detail here. A laser (depicted in blue), which is splitted at a 50/50 beam splitter arrive at the same time at crystal 1 and 2. For that, the pathlength between the beam splitter and the crystals 1 and 2 (written as $\ell_{p1}$ and $\ell_{p2}$) need are matched with a standard trombone system.

Crystals 1 and 2 can both produce photon pairs with horizontal polarization, while crystal 3 and 4 can both produce vertically polarized photon pairs -- with non-colinear phase-matching conditions. This can be achieved, in analogy to the cross-crystal two-photon source, with type-I SPDC and a diagonally polarized pump beam.

In order to make sure that the states are produced in a coherent superposition, path lengths need to be chosen appropriately. In the example above (Figure \ref{figSI:4GHZ}), it needs to be fulfilled that one cannot distinguish whether the 4-fold coincidence count came from crystal 1 and 2, or whether it came from crystal 3 and 4. For that, the path lengths $\ell_1$, $\ell_2$, $\ell_3$ and $\ell_4$ need to be equal within the coherence length of the down-converted photons. Furthermore, the pump laser paths need to be matched such that $\ell_{p_1}=\ell_{p_2}$ and $\ell_{p_3}=\ell_{p_4}$ within the coherence time of the laser. The final requirement is that the path lengths of the laser between the two crystals must be matched with the path length of the down-converted photons, such that $\ell_{p_3}=\ell_1$ within the coherence time of the laser. It can be summarized as
\begin{subequations}
\begin{align}
&|\ell_i - \ell_j| \ll c\tau_{SPDC},\\
&|\ell_{p1} - \ell_{p2} | \ll c\tau_{pump}, |\ell_{p3} - \ell_{p4} | \ll c\tau_{pump},\\
&|\ell_{p3} - \ell_i | \ll c\tau_{pump}, |\ell_{p4} - \ell_i | \ll c\tau_{pump}, 
\label{coherenceRequ}
\end{align}
\end{subequations}   
where $c\tau_{SPDC}$ and $c\tau_{pump}$ are the coherence length of the down-converted photons and the pump laser, respectively. We find these restrictions on the temporal coherence and indistinguishability by applying the results that were obtained for the two-photon case by Jha et al. \cite{jha2008temporal}. These are similar requirements as for standard 4-photon GHZ sources based overlapping the photons at a polarizing beam splitter. Furthermore, also the requirements for the matching of frequencies is analogous to the standard 4-photon GHZ source using an PBS.

For other polarization experiments, very similar requirements are necessary. In experiments involving spatial modes (such as in Figure \ref{fig:4partyhighdim} or Figure \ref{figSI:n6d5GHZ} in the Appendix), it is further important to eliminate spatially dependent phase shifts acquired on propagation of pump and down-conversion photons. This can be done with using standard 4-f lens systems.

We show in Section II that if there are misalignments between the overlapping paths, it does not reduce the coherence between the different terms but changes the relative amplitudes between them. This is because misaligned beams do not arrive at the detectors and consequently do not lead to a four-fold coincidence count. Very similar conditions apply for all other experimental setups presented in the main text.

\section{Appendix II. Detailed calculation of the quantum states and the effect of misalignment}
Here we perform the detailed calculation of the 4-photon GHZ experiment. The down-conversion process can be described as a series expansion in the form of 
\begin{align}
\hat{U}^{P_1P_2}_{a,b}=1&+g\left(\hat{a}_{a,P_1}^{\dagger} \hat{a}_{b,P_2}^{\dagger} - \hat{a}_{a,P_1} \hat{a}_{b,P_2}\right) \nonumber\\
&+ \frac{g^2}{2} \left(\hat{a}_{a,P_1}^{\dagger} \hat{a}_{b,P_2}^{\dagger} - \hat{a}_{a,P_1} \hat{a}_{b,P_2}\right)^2 + \mathcal{O}(g^3)
\label{4GHZfull}
\end{align}
where $\hat{a}_{a,P}^{\dagger}$ and $\hat{a}_{a,P}$ are creation and annihilation operators for a photon in the mode $a$ and with polarization $P$, respectively, and $g$ is proportional to the SPDC rate and the pump power. For simplicity, we restrict ourselves to single-mode analysis. In the 4-photon GHZ setup, four crystals are used, therefore the state can be expressed as
\begin{align}
\ket{\psi}=\hat{U}^{VV}_{c,d}\hat{U}^{VV}_{a,b}\hat{U}^{HH}_{b,d}\hat{U}^{HH}_{a,c}\ket{vac},
\label{4GHZpsi}
\end{align}
where $\ket{vac}$ is the vacuum state. In addition, we want to consider the effect of misalignment between the modes from the different crystals. As introduced in \cite{zou1991induced}, if the down-converted modes are perfectly overlapped, they can be treated as one mode. In the extreme case of complete misalignment, the photons do not reach the detector, which can be described by inserting a beam stop between the crystals. For the intermediate cases, we can describe the situation using a variable beam splitter with transmissivity T and reflectivity R:
\begin{align}
\mathcal{\bar{T}}_a(\hat{a}_{a,P}^{\dagger}) = T_a \hat{a}_{a,P}^{\dagger} + R_a \hat{a}_{0,P}^{\dagger}
\label{4GHZT}
\end{align}
where $\hat{a}_{0,P}^{\dagger}$ is an empty loss mode, and $|R|^2+|T|^2=1$. Considering misalignment in all four arms, which we denote as $\mathcal{T}_{(a,b,..,n)}(x)=\mathcal{\bar{T}}_a(\mathcal{\bar{T}}_b(..(\mathcal{\bar{T}}_{n}(x)..))$, we find
\begin{align}
\ket{\psi}&=\hat{U}^{VV}_{c,d}\hat{U}^{VV}_{a,b}\mathcal{T}_{(a,b,c,d)}\left(\hat{U}^{HH}_{b,d}\hat{U}^{HH}_{a,c}\right)\ket{vac}\nonumber\\
&=g^2 \Big(\hat{a}_{a,V}^{\dagger} \hat{a}_{b,V}^{\dagger} \hat{a}_{c,V}^{\dagger} \hat{a}_{d,V}^{\dagger} \nonumber\\
&+ T_a T_b T_c T_d \hat{a}_{a,H}^{\dagger} \hat{a}_{b,H}^{\dagger} \hat{a}_{c,H}^{\dagger} \hat{a}_{d,H}^{\dagger}\Big)\ket{vac}\nonumber\\
&+\frac{g^2}{2}\Big(T_a T_c\hat{a}_{a,H}^{\dagger 2} \hat{a}_{c,H}^{\dagger 2} + T_b T_d \hat{a}_{b,H}^{\dagger 2} \hat{a}_{d,H}^{\dagger 2} \nonumber\\
&+ \hat{a}_{a,V}^{\dagger 2} \hat{a}_{b,V}^{\dagger 2} + \hat{a}_{c,V}^{\dagger 2} \hat{a}_{d,V}^{\dagger 2}\Big)\ket{vac}\nonumber\\
&+(\textnormal{less than four photons}) + \mathcal{O}(g^3)
\label{4GHZpsi2}
\end{align}
If $T_i=1$, the 4-photon terms of equation (\ref{4GHZpsi2}) are represented by the first line of equation (\ref{4partyGHZ}).

Terms that do not lead to four photons in the spatial modes $a$, $b$, $c$, $d$ are discarded by the four-fold coincidence detection, and higher order terms in $g$ are discarded as their probability is small for $g\ll 1$. Therefore we find that the resulting state can be written as
\begin{align}
\ket{\psi}&=\frac{\ket{V_a,V_b,V_c,V_d} + T_a T_b T_c T_d \ket{H_a,H_b,H_c,H_d}}{\sqrt{1+T_a^2 T_b^2 T_c^2 T_d^2}}
\label{4GHZpsi3}
\end{align}
That means that the coherence between the state does not decrease, but the state is not equally distributed thus it is not maximally entangled. However, reducing the laser power before crystal 3 and 4 (thus changing the value of $g$ for the crystal 3 and 4) can compensate for the drop in entanglement. Conversely, a filter in the down-converted photons could be used to compensate for lower pump power. If $T_i=1$, equation (\ref{4GHZpsi3}) is the same as the second line of equation (\ref{4partyGHZ}). In the following examples, for simplicity we set $T_a=T_b=..=T$.

For a six-photon GHZ state in polarization, an analogous calculation with
\begin{align}
\ket{\psi}&=\hat{U}^{VV}_{a,b}\hat{U}^{VV}_{c,d}\hat{U}^{VV}_{e,f}\mathcal{T}_{(a,b,c,d,e,f)}\left(\hat{U}^{HH}_{a,c}\hat{U}^{HH}_{b,e}\hat{U}^{HH}_{d,f}\right)\ket{vac},
\label{6GHZpsi1}
\end{align}
leads to the six-photon GHZ state 
\begin{align}
\ket{\psi}&=\frac{\ket{V_a,V_b,V_c,V_d,V_e,V_f} + T^6 \ket{H_a,H_b,H_c,H_d,H_e,H_f}}{\sqrt{1+T^{12}}}.
\label{6GHZpsi3}
\end{align}
Similarly, for the W-state, the experiment can be described by
\begin{align}
\ket{\psi}=&\hat{U}^{HV}_{a,b}\hat{U}^{HH}_{c,d}\mathcal{T}_{(a,b,c,d)}\big(\nonumber\\
&\hat{U}^{VH}_{a,b}\hat{U}^{HV}_{a,c}\hat{U}^{HH}_{b,d}\hat{U}^{HV}_{a,d}\hat{U}^{HH}_{c,b}\big)\ket{vac}
\label{4Wpsi1}
\end{align}
which leads to (accounting for misalignment between each crystal, and using four-fold coincidences) to
\begin{align}
\ket{\psi}&=\frac{1}{N}\Big(T^4\ket{H_a,H_b,H_c,V_d} + T^4 \ket{H_a,H_b,V_c,H_d}\nonumber\\
&+ T^2 \ket{H_a,V_b,H_c,H_d} + \ket{V_a,H_b,H_c,H_d}\Big),
\label{4Wpsi2}
\end{align}
where $N$ is a normalization constant. In this experiment, we expect that the noise from higher-order terms is slightly higher because of induced emission due to the fact that the input states are the same as the states produced in the crystals (in contrast to other examples). The different coefficients in front of the terms can be compensated again by adjusting the power of the pump laser.

The examples considering a high-dimensional degree of freedom (such as orbital angular momentum (OAM) of photons) use mode shifters in addition to down-conversion crystals (figure \ref{fig:4partyhighdim}a in the main text), which can be defined as
\begin{align}
\mathcal{\bar{S}}_a\left(\hat{a}_{a,\ell}^{\dagger}\right) = \hat{a}_{a,\ell+1}^{\dagger}
\label{ModeShift}
\end{align}
and $\mathcal{S}_{(a,b,..,n)}(x)=\mathcal{\bar{S}}_a(\mathcal{\bar{S}}_b(..(\mathcal{\bar{S}}_{n}(x)..))$. Then, the state in the four-photon three-dimensional experiment can be described as
\begin{align}
\ket{\psi}&=\hat{U}_{a,b}\hat{U}_{c,d}\mathcal{T}_{(a,b,c,d)}\big(\nonumber\\
&\mathcal{S}_{(a,b,c,d)}\big(\hat{U}_{a,c}\hat{U}_{b,d}\mathcal{S}_{(a,b,c,d)}\big(\hat{U}_{a,d}\hat{U}_{b,c}\big)\big)\big)\ket{vac}
\label{4GHZ3dim1}
\end{align}
where $\hat{U}$ produces the same polarization, and zero-order OAM modes. A completely analogous calculation as above leads to
\begin{align}
\ket{\psi}=\frac{1}{N}\Big(\ket{0_a,0_b,0_c,0_d}&+T^4\ket{1_a,1_b,1_c,1_d}\nonumber\\
&+T^4\ket{2_a,2_b,2_c,2_d}\Big),
\label{4GHZ3dim2}
\end{align}
where again the different coefficients, which occur due to imperfect overlapping of the modes, can be compensated by adjusting the pump power in crystal 5 and 6.

Finally, the experiment for the two-photon high-dimensional entangled state allows for adjustable phases, which can be written as
\begin{align}
\mathcal{P}_{a,\phi}\hat{a}_{a,\ell}^{\dagger} = \exp\big(i \phi\big) \hat{a}_{a,\ell}^{\dagger}
\label{ModeShift}
\end{align} 
For example, with four crystals in a row with mode shifters constantly adding +1, and phase shifters changing the phase by $\pi/2$, we can write
\begin{align}
\ket{\psi}&=\hat{U}_{a,b}\mathcal{T}_{a}\big(\mathcal{S}_{(a,b)}\mathcal{P}_{b,\frac{\pi}{2}}\big(\hat{U}_{a,b}\mathcal{S}_{(a,b)}\mathcal{P}_{b,\frac{\pi}{2}}\big(\hat{U}_{a,b}\nonumber\\
&\mathcal{S}_{(a,b)}\mathcal{P}_{b,\frac{\pi}{2}}\big(\hat{U}_{a,b}\mathcal{S}_{(a,b)}\mathcal{P}_{b,\frac{\pi}{2}}\hat{U}_{a,b}\big)\big)\big)\big)\ket{vac}.
\label{2particleHighdim1}
\end{align}
This leads to
\begin{align}
\ket{\psi}=\frac{1}{N}\Big(\ket{0_a,0_b}+iT\ket{1_a,1_b}&-T\ket{2_a,2_b}-iT\ket{3_a,3_b}\Big)
\label{2particleHighdim2}
\end{align}
Similarly as in examples above, misalignment reduces the entanglement by unweighting the state. However, that effect can be compensated by adjusting the pump power before each crystal. Therefore maximally entangled, arbitrary, high-dimensional entangled two-photon states can be created.  
\section{Appendix III. Construction of general experiments}\label{app3}
\textit{Two-Photon Case} -- For two photons, arbitrary high-dimensional quantum states can be created. If d crystals are pumped coherently as shown in Figure 4b in the main text, the resulting state is in a superposition of being created in either of the crystals. As the down-converted photons can be manipulated between each crystal, the mode-number and phase can be adjusted for each individual term of the complete state.

\begin{figure}[ht!]
\includegraphics[width=0.5 \textwidth]{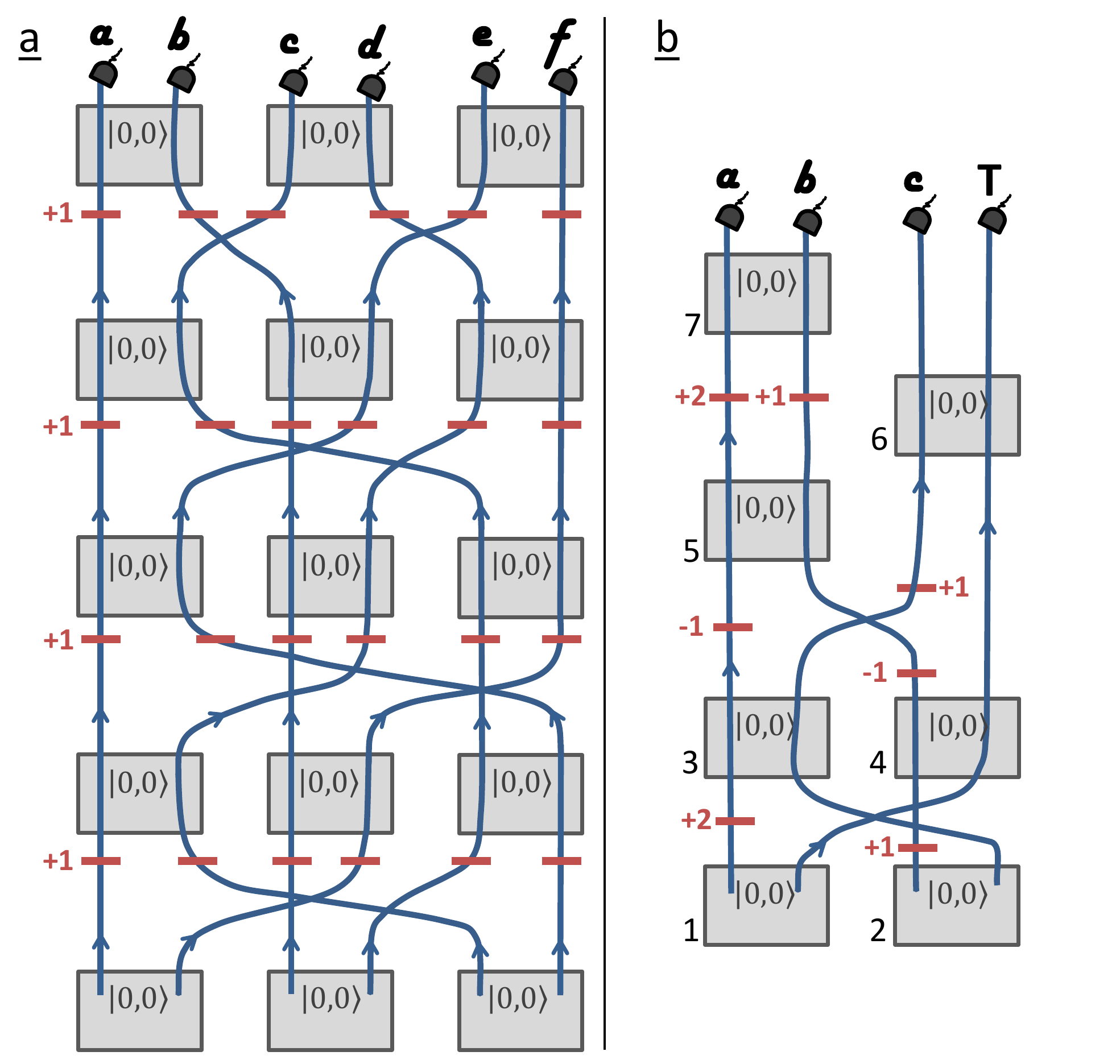}
\caption{\textbf{a}: The experimental configuration for a 6-particle 5-dimensional GHZ state. \textbf{b}: An experiment which creates an asymmetricly entangled quantum state, with a Schmidt-Rank-Vector of (4,2,2). One photon (in detector T) is used to trigger the three-photon state in (a,b,c).}  
\label{figSI:n6d5GHZ}
\end{figure}

\textit{Multi-Photon Case} -- We analyse the case where n-photon states are created in n different paths. That requires at least $c=\frac{n}{2}$ crystals to fire simultaneously, and together emit photons in n different paths. In the example \ref{figSI:n6d5GHZ}a, $n=6$ and $c=3$ (six photons are created with three crystals - which is represented in every row of the setup). There are $d=n-1$ different ways to arrange $c$ crystals such that they produce n-fold coincidence counts. In that example, $d=5$ and a n=6-fold coincidence count can be created by $c=3$ crystals emitting in six paths (ab-cd-ef, ac-be-df, ad-bf-ce, ae-bd-cf, af-bc-de).
\begin{figure}[ht!]
\includegraphics[width=0.5 \textwidth]{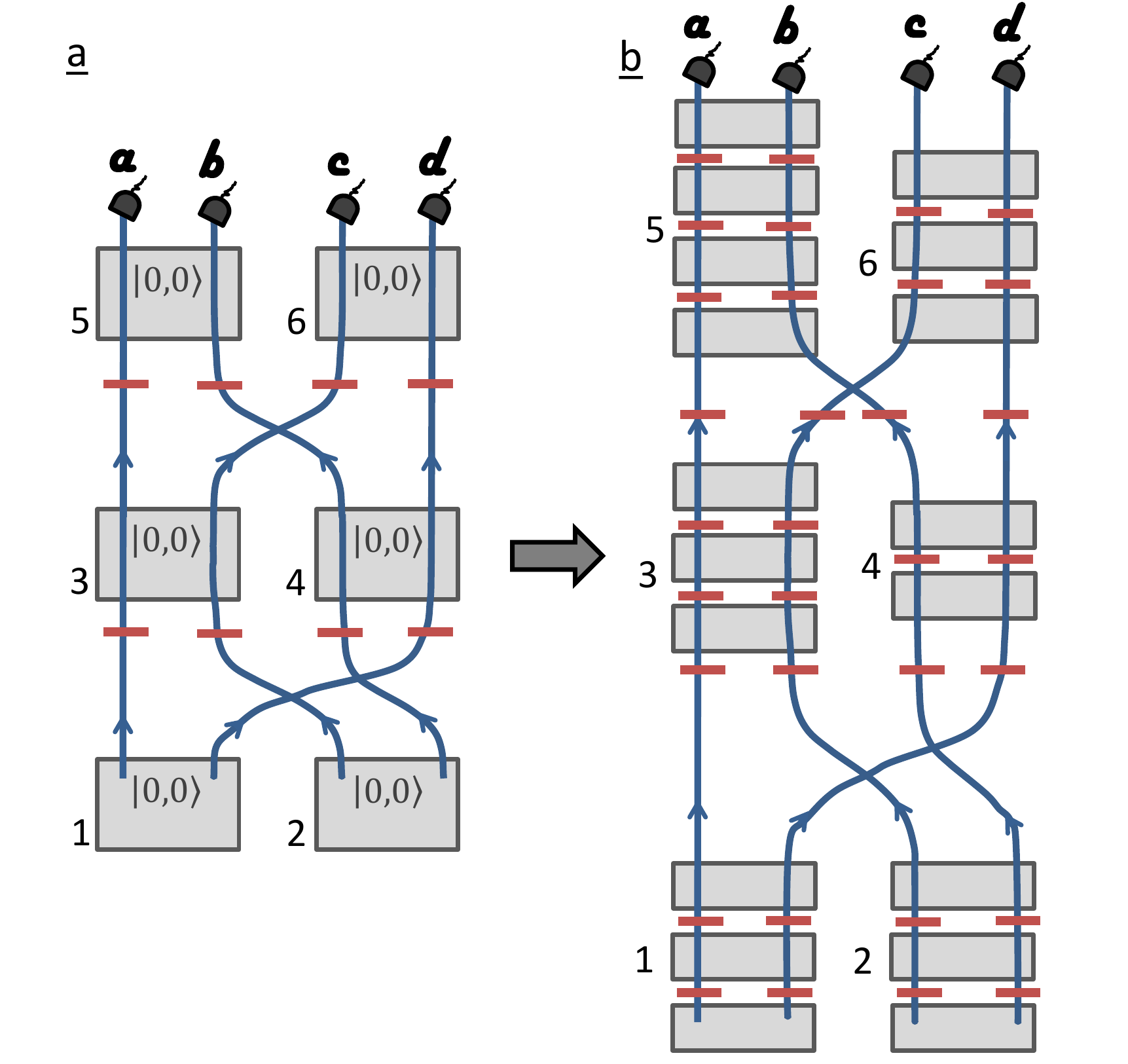}
\caption{\textbf{a}: Experiment for an arbitrary 4-photon 3-dimensional GHZ state \textbf{b}: A generalisation of GHZ state. Every crystal can be replaced by a row of crystals each producing arbitrary high-dimensional 2-photon states. The red lines indicate an arbitrary mode shifter (such as holograms for OAM).}  
\label{figSI:nGeneral}
\end{figure}
That means, arbitrary (n-1)-dimensionally entangled n-photon GHZ state can be created. For the case of n=4, that can be written as
\begin{align}
\ket{\psi_1}=\frac{1}{N}\big(&\ket{\ell_1, \ell_2, \ell_3, \ell_4}+\nonumber\\
&c_{1,1} \ket{\ell_5, \ell_6, \ell_7, \ell_8}+\nonumber\\
&c_{1,2} \ket{\ell_9, \ell_{10}, \ell_{11}, \ell_{12}}\big)
\end{align}
where $\ell_i \in \mathbb{Z}$ and $c_{i,j} \in \{0,e^{i\phi}\}$ (which is depicted in Figure \ref{figSI:nGeneral}a).

Each crystal could now be replaced by a row of crystals (such as Figure 4b in the main text), where each row produces an arbitrary 2-photon d-dimensional entangled states. That leads to the following possible state
\begin{align}
&\ket{\psi_2}=\frac{1}{N}\Big(\nonumber\\
&\big( \sum_i c_{1,i} \ket{\ell_{1,i}, \ell_{2,i}}_{a,b}\big) \big( \sum_i c_{2,i} \ket{\ell_{3,i}, \ell_{4,i}}_{c,d}\big)+\nonumber\\
&\big( \sum_i c_{3,i} \ket{\ell_{5,i}, \ell_{6,i}}_{a,c}\big) \big( \sum_i c_{4,i} \ket{\ell_{7,i}, \ell_{8,i}}_{b,d}\big)+\nonumber\\
&\big( \sum_i c_{5,i} \ket{\ell_{9,i}, \ell_{10,i}}_{a,d}\big) \big( \sum_i c_{6,i} \ket{\ell_{11,i}, \ell_{12,i}}_{b,c}\big)\Big)
\end{align}
with and $\ell_{i,j} \in \mathbb{Z}$. Furthermore, one can use filters both in the pump and down-converted photon paths, such that $c_{i,j} \in \mathbb{C}$ with $|c_{i,j}|\leq 1$. 
The W-state (presented in figure 3 of the main text) is a special case of figure \ref{figSI:nGeneral}. A high-dimensional expample is shown in figure \ref{figSI:n6d5GHZ}b. It shows a setup for creating n=3 photon entangled state (with one photon acting as a trigger $\ket{0_T}$) with $d=4$ terms in an asymmetric configuration ($\ket{\psi}=\frac{1}{2}\ket{0_T}\left(\ket{0_a0_b0_c}+\ket{1_a0_b1_c}+\ket{2_a1_b0_c}+\ket{3_a1_b1_c}\right)$). The state can be quantified by the Schmidt-Rank Vector (4,2,2) \cite{huber2013structure}. A variaty of similar states can be written down in that form. 
\vspace{25px}

\section{Appendix IV. Efficiency of state creation}\label{app4}
The method is probabilistically (not every generated n-photon state leads to a valid n-fold photon state with one photon in each of the n paths). The effciency E is the number of valid n-fold events divided by the number of possible n-photon events. For d-dimensional n-particle GHZ states, there are d combinations that lead to a valid n-fold, and for c crystals there are c$^{n/2}$ possible combinations leading to a n-photon states. For GHZ-states, $c=\frac{n\cdot d}{2}$. This leads to an effciency of $E = \frac{d}{\left(\frac{n\cdot d}{2}\right)^{n/2}}$.

\section{Appendix V. How we found the technique}\label{app5}
\begin{figure}[ht!]
\includegraphics[width=0.5 \textwidth]{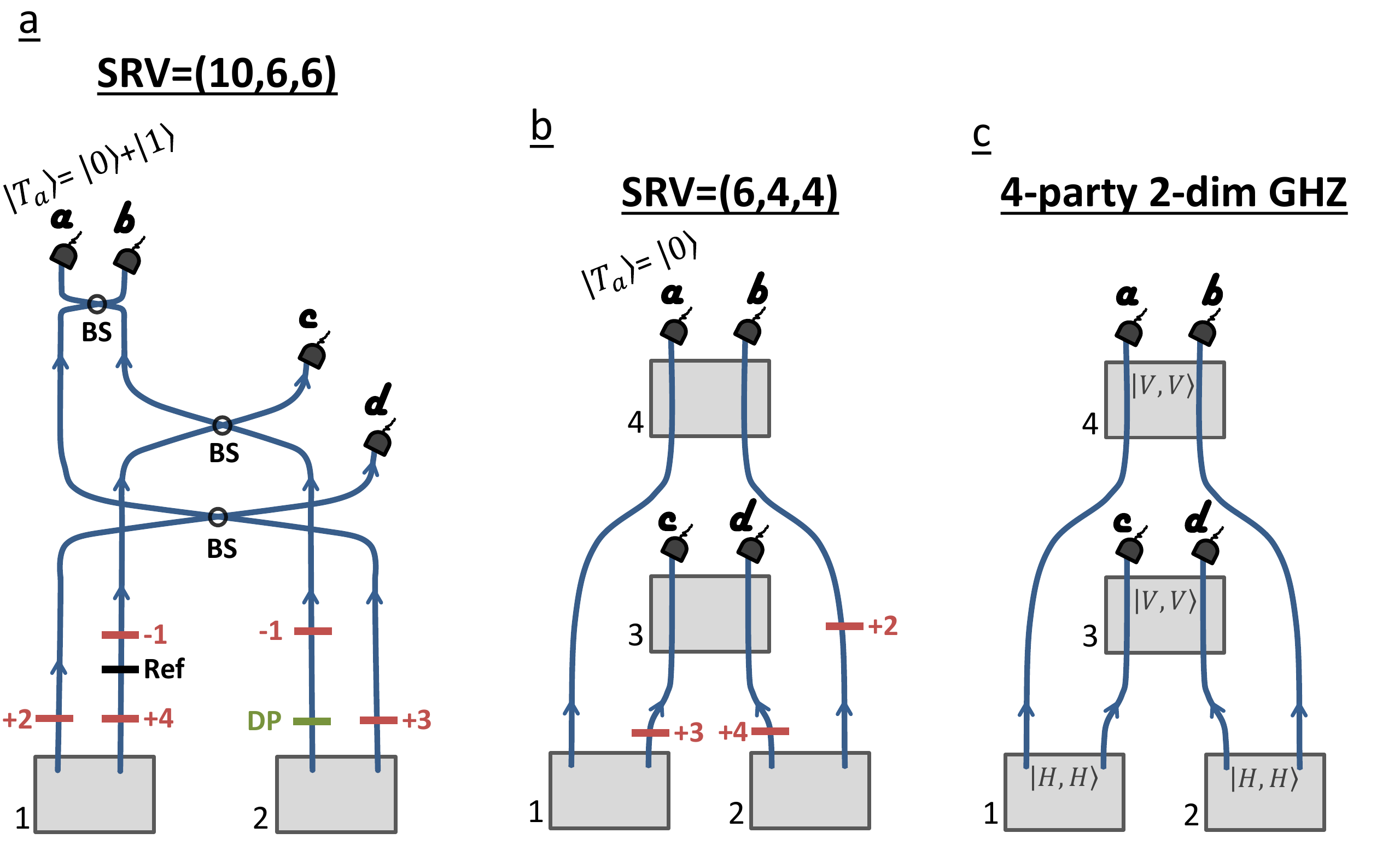}
\caption{\textbf{a}: We investigated an experiment with an unexpectedly large Schmidt-Rank Vector (10,6,6) \cite{huber2013structure}. Each crystal produces a three-dimensional two-photon entangled state, and $\ket{T_a}$ stands for the trigger in mode $a$. The experiment works because the crystals are producing four-photons in a coherent superposition of one SPDC event in each of the crystals simultaneously, and two SPDC events in each of the crystals. \textbf{b}: By explicitly allowing superpositions of crystals and enableing overlapping of their output paths (but removing any other interaction such as beam splitters), the program was still able to find high-dimensional multi-photon experimental setups. \textbf{c}: Restricting the program further to allow only for two-dimensional polarisation and no trigger, it still found a solution. From that solution we understood the very basic idea, which we were able to generalize further.}  
\label{fig:HowWeFoundIt}
\end{figure}

The computer algorithm described in \cite{krenn2016automated} produced setups for experimentally feasible high-dimensional multi-photon states. We analyzed a result of an unusually high-dimensional entangled 3-photon state, as it was not expected that such a state can be created, given the restricted set of available elements of the algorithm (figure \ref{fig:HowWeFoundIt}a). One restriction was that each setup starts with two SPDC crystals, which can produce each 3-dimensional 2-photon entanglement. However, in that particularly unusual solution, the algorithm used three types of available SPDC process in a superposition (crystal 1 and crystal fires twice, and crystal 1 and 2 fire once together). Upon understanding that technique, we gave the algorithm explicitly the possibility to use several crystals in superposition and overlapping their paths. We further restricted other available elements, in order to understand how powerful this technique is. The algorithm was able to find a high-dimensional entangled state using only crystals and mode shifters (figure \ref{fig:HowWeFoundIt}b). We further restricted it to the simplest case of 2-dimensional polarization, were it also found solutions (figure \ref{fig:HowWeFoundIt}c). From there on, the we extracted the principle idea of the technique and subsequently generalized it.

\end{document}